# XDC Network Assessment: Decentralization, Scalability and Security

Technical Whitepaper

Mohuya Chakraborty (mohuyacb@gmail.com)

Atul Khekade (atul@xinfin.org)

**Abstract.** XinFin, in 2019, unveiled the XDC network, an enterprise-ready hybrid blockchain platform that is open-source and specializes in tokenization for real-world decentralized finance. Overseeing the XDC network is currently the XDC Foundation, a non-profit organization established to encourage the growth, enhancement, and adoption of the XDC Network through community-driven projects such as GitHub. This whitepaper discusses the real-time assessment of the XDC network's decentralization, scalability, and security aspects as well as the Nakamoto coefficient estimation that follows—a measure of a decentralized system's decentralization nature that quantifies the minimal number of nodes or entities needed to compromise the system. A high coefficient denotes greater decentralization, while a low number denotes increased disruption risk. The XDC network's real-time computation of the high Nakamoto coefficient demonstrates its highly decentralized character. The article also addresses the diversity of consensus and execution clients, the host distribution, the geo-distribution, and some of the outstanding issues and business considerations.

## 1 Introduction

To guarantee the transparency, dependability, and security of the digital assets under investigation, this whitepaper has been crafted to act as the foundation for assessing the XDC blockchain network's scalability, security, functionality, and general integrity. The blockchain trilemma tackles the difficulties programmers encounter in building a decentralized, scalable, and secure blockchain without sacrificing any of these qualities. Decentralization is the process of building a blockchain system that does not require control from a central point of contract; scalability is the capacity to handle a growing volume of transactions; and security is the ability of the blockchain system to function as intended and protect itself against attacks, bugs, and other unforeseen issues. Instant block finality, hybrid relay bridges, ISO 20022 financial messaging standards, and an open-source, global XinFin Delegated Proof of Stake consensus network (XDPoS) are all supported by the blockchain solution known as the XDC network [1], created by XinFin [2]. The network's hybrid (private/public) architecture aims to make asset tokenization and institutional trade finance applications easier. XDC is the fundamental utility token of the XDC Network. Its hybrid design additionally enables private transactions over private subnetworks for the purpose of securing sensitive data. The network's ISO 20022 compliant messaging system ensures interoperability with Enterprise Resource Planning (ERP), the Society for Worldwide Interbank Financial Telecommunication (SWIFT), and core banking systems [3]. Cross-chain Ethereum-based tokens and projects, including ERC-20 and ERC721 [4], can be migrated to the network, which offers nearly non-existent gas costs and significantly improves scalability and speed. It makes sense for the XDC Network to act as a conduit between public blockchain networks and legacy systems. Through three main points—the network, developer ecosystem, and business prospects—the paper explores the qualitative and technical evaluation of the XDC network, covering its governance model, consensus mechanism, scalability, interoperability, developer environment, and market viability.

The rest of the paper is organized as follows. Section 2 addresses the qualitative and technical assessment through three key indicators. Section 3 examines the XDC network through its technical and security aspects focussing on the decentralization capability, validator participation, and client distribution. The most important contribution to this section is the computation of the Nakamoto coefficient that measures the nature of decentralization of the network. Section 4 focusses on the developer ecosystem and assesses the vibrancy and activities of the developer community such as GitHub and leverages its statistics to gauge community engagement and project health. Section 5 is all about evaluating the practical business implications of using the XDC network including transaction costs and predictability for understanding its economic viability. Section 6 presents some of the open issues and challenges. Section 7 concludes the paper with some concluding remarks and highlights on future prospects.

## 2   Qualitative and Technical Assessment

The primary metrics used for evaluating the XDC network's technical and qualitative qualities are presented in this section. These metrics have been meticulously selected to offer a thorough assessment that addresses a range of factors, including developer involvement, security, network decentralization, and economic viability. Metrics include anecdotal insights from GitHub statistics as well as quantifiable data points like the Nakamoto coefficient. When taken as a whole, these metrics provide stakeholders with a comprehensive understanding of the advantages and disadvantages of the XDC network, facilitating wise decision-making. To facilitate a structured and focused analysis, the key metrics have been categorized into three main groups:

1. *Network* - The "Network" category explores the security and technological components of the XDC network, observing things like client distribution, validator participation, and decentralization.
2. *Developer Ecosystem* - Using GitHub information primarily, the number of repositories, open issues, network, and subscriber counts are analyzed by the

"Developer Ecosystem" group to gauge the vitality and activity of the developer community.
3. *Business* - This category assesses the real-world effects of utilizing the XDC network, taking into account variables like transaction costs and their predictability.

# 3 Network

This section examines the XDC network with regard to its decentralization capability, participation rate, number of validators, and client distribution.

## 3.1 Network Decentralization

In a blockchain network, decentralization is the transfer of power and decision-making from a centralized body to a more expansive, dispersed network. There are several advantages and capabilities of blockchain decentralization. Decentralization offers enhanced security by spreading data among nodes, making the network more resistant to fraud and threats. It also provides greater transparency through a public ledger, ensuring data reliability and unchangeability. Decentralized networks create a trustless environment, allowing for better data reconciliation and efficient resource allocation. They also enable real-time, shared data views, reducing the risk of data loss or corruption. Decentralization also reduces the possibility of bottlenecks and systemic failures by not relying on a single entity [5-6].

*Nakamoto Coefficient:* The most important metric to quantify the level of decentralization of any blockchain network is the Nakamoto Coefficient. A higher Nakamoto Coefficient indicates a more decentralized network, reducing the risk of single points of failure or control, whereas a lower Nakamoto Coefficient indicates associated risks of centralization. In the context of network security, this metric is used to evaluate how robust the network is against potential attacks. A higher coefficient suggests that the network is more secure and resistant to threats like 51% attacks. The Nakamoto Coefficient measures the minimum number of validators or nodes required to compromise a blockchain network with 51% attack i.e., it represents the minimum number of entities (nodes or validators) that control more than 50% of the total power of the network. It is a useful metric for understanding the decentralization of a blockchain. A higher coefficient indicates greater decentralization, meaning more nodes would be needed to collude to compromise the network [7].

Since XDC is a Delegated Proof of Stake (DPoS) network, there comes the need to examine the distribution of staked tokens among validators and identify the point at which a group of entities controls more than 33.33% of the network. The Nakamoto Coefficient is the number of entities required to reach this threshold [8]. In the XDC network, there are 108 validator masternodes and 198 standby nodes. Validator masternodes operate and participate in XDC Network's XDPoS consensus engine, validating transactions and block creation. Standby nodes are identical in form and function to validators but do not participate in validating transactions and block creation. However, they must have same stake as that of the validators masternodes i.e., 10 million XDC. So, in short, each of the validator masternodes and standby nodes should have a stake of 10 million XDC to participate in the network's consensus.

The Nakamoto Coefficient for the XDC network computation as on 19-07-2024 at 15:08 IST is presented in Table 1.

Table 1 Nakamoto Coefficient

| Total Stakeholder Nodes | Total Stakes | 33.33% Stakes | Number of Nodes to reach Threshold | Nakamoto Coefficient |
|---|---|---|---|---|
| 108 (Validators) + 198 (Standby) = 306 | 306 x 10 million = 3060 million XDC | 1019.898 ~ 1020 million | 1020/10 = 102 | 102 + 1 = 103 |

This high value of Nakamoto Coefficient of the XDC network shows its effectiveness in maintaining a very high decentralization level as compared to other blockchain networks [8].

### 3.2 Participation Rate and Number of Validators

*Participation Rate:* This is calculated as the percentage of active validators in relation to the total number of validators. A higher participation rate generally indicates a healthier and more active network.
*Number of Validators:* This represents the total count of validators that are actively participating in the consensus mechanism of the network. A higher number can indicate a more decentralized and secure network.

The details of the specifications as on 19-07-2024 at 15:08 IST are depicted in Table 2 [9-10].

Table 2 Participation Rate and Validators

| Total Nodes | Active Nodes | Participation Rate | Number of Validators | Number of Standby Nodes | Total Stakeholder Nodes | Resigned Nodes | Slashed Nodes |
|---|---|---|---|---|---|---|---|
| 430 | 215 | 50% | 108 | 198 | 306 | 10 | 11 |

### 3.3 Client Distribution

*Geo Distribution:* Geo distribution refers to the physical locations of the nodes that participate in the network. These nodes can be spread across continents, countries, and cities. The geographical spread of nodes is crucial for several reasons. A geographically dispersed network offers resilience to localized events, reduced latency for transactions and smart contract executions, and regulatory diversification, as it can handle different jurisdictions' regulations. This ensures the network remains operational even in disruptions, reducing the risk of being heavily impacted by regulatory changes in any single jurisdiction. Real-time data as on 19-07-2024 at 15:08 IST shows 20 countries actively participating as execution and consensus clients [9]. Table 3 shows the node distribution of top 10 countries.

Table 3  Geo Distribution of XDC Clients (Top 10 Countries)

| Sl. No. | Country | Number of Nodes |
|---|---|---|
| 1 | Singapore | 127 |
| 2 | Germany | 114 |
| 3 | United States | 90 |
| 4 | United Kingdom | 28 |
| 5 | Netherlands | 16 |
| 6 | France | 10 |
| 7 | Australia | 10 |
| 8 | Canada | 10 |
| 9 | Finland | 8 |
| 10 | Japan | 6 |

*Host Distribution:* Host distribution examines the variety of hosting environments for the nodes of the network. This includes cloud services (like AWS, Google Cloud, Azure), dedicated hosting facilities, and private servers. A diverse host distribution is beneficial for several reasons. A mix of hosting options can help mitigate platform-specific risks, avoid centralization, and optimize overall network costs. Relying on a single cloud provider exposes the network to outages or disruptions, while avoiding centralized control can prevent undue influence. A mix of hosting options ensures resilience and cost optimization. Table 4 presents the host distribution of clients of the XDC network.

Table 4  Host Distribution of XDC Clients

| Sl. No. | Type of Host | Type of Engagement |
|---|---|---|
| 1 | Enterprises and Corporations | Many nodes are hosted by enterprises and corporations that utilize the XDC Network for its enterprise-grade features, such as enhanced security, scalability, and interoperability. These organizations can range from financial institutions to supply chain companies that provide high-performance XDC Network nodes and APIs, enabling businesses to host their XDC nodes for real-time settlements and improved business functions [11], and cloud service providers that are recommended for hosting XDC Masternodes due to their reliable infrastructure [12]. |
| 2 | Independent Validators | Blockchain enthusiasts, small businesses, or tech-savvy individuals who participate in the network's consensus process. However, they need to comply with KYC (Know Your Customer) requirements to maintain transparency and security [13]. |
| 3 | Cloud Service Providers | A significant number of nodes are hosted on cloud existing platforms. Cloud hosting offers scalability, reliability, and ease of management for running consensus nodes [12]. |
| 4 | Data Centers | Hosting of some nodes to provide robust infrastructure and security measures for maintaining the network's integrity and performance. These data centers are typically tier III and IV, ensuring high reliability and performance. For example, in Australia, nodes are hosted in major cities like Sydney, Melbourne, Brisbane, and Perth [13]. |
| 5 | Educational and Research Institutions | Some nodes may be hosted by universities or research institutions as part of blockchain research data management, decentralized learning platforms and credential verification. |
| 6 | Government and Public Sector | Running nodes for<br>i) Compatibility with legacy systems using ISO20022 financial messaging standard for seamless integration with traditional financial systems [14].<br>ii) Transparency and security for projects requiring high level of accountability [15]. |
| | | iii) Compatibility with trade finance that supports dApps facilitating secure and efficient trade finance operations [16].<br>iv) Cost effectiveness with transaction costs significantly lower than those of other blockchains [15].<br>v) Real-world projects like launch of regulated security tokens and stablecoins that have implications for public sector financial management [16]. |
| 7 | Blockchain and Fintech Startups | Runing nodes to develop, test, and deploy their applications on the XDC Network, contributing to the overall node distribution. |

*Client Diversity:* Client diversity refers to the variety of software clients used to run nodes on the network. Client diversity is crucial for several reasons. Different clients have different codebases, reducing the risk of network-wide failure or attack. This encourages innovation and competition, as teams collaborate to enhance performance and features. Operators can choose clients that fit their needs, preferences, or hardware configurations, promoting a more inclusive network.

*Client Types:* In any blockchain network, there are two types of clients: execution and consensus. In the XDC Network, the roles of execution clients and consensus clients are different but balancing.

- *Execution Clients*: They manage the computational tasks. Their role is to handle the processing of transactions, execution of smart contracts, and maintenance of the blockchain state. The functionality of these clients is to listen to new transactions, execute them, and update the state of the blockchain. This includes managing the database of all blockchain data. They are responsible for ensuring that all transactions and smart contracts are processed correctly. They are also called full nodes while maintaining the state of the blockchain and providing transaction history.
- *Consensus Clients:* They ensure agreement across the network. Their role is to ensure that all nodes in the network agree on the state of the blockchain. They manage the consensus mechanism, which in the case of the XDC Network is XDPoS. The functionalities of these nodes include validation of the data from execution clients and ensuring that all nodes reach a consensus on the blockchain's state. This involves selecting and validating blocks proposed by the masternodes. In the XDC Network, consensus clients work with the 108 masternodes to maintain network security and efficiency. The consensus clients are also called masternodes. These are the primary consensus clients in the XDC Network. They participate in the XDPoS consensus mechanism. There are two types of masternodes viz., Validator Masternodes and Standby Masternodes. The former nodes actively participate in validating transactions and block creation whereas the latter nodes are ready to take over if a validator masternode drops out.

The exact number of these nodes can vary as the network evolves and more participants join or leave. The XDC Network aims to maintain a robust and scalable infrastructure to support its operations. All the masternodes are full nodes. They store a copy of the blockchain, produces blocks and keeps the chain consistent. These nodes are controlled by consortium members and come with a number of caveats. But not all full nodes are masternodes. Masternodes have additional responsibilities and privileges within the network. The details of the specifications as on 19-07-2024 at 15:08 IST are presented in Table 5 [9].

Table 5  Client Types

| Total Nodes | Consensus Nodes | Execution Nodes |
|---|---|---|
| 430 | 108 (Validator) + 198 (Standby) = 306 | 430 – 306 = 124 |

## 4  Developer Ecosystem

The XDC Network uses several metrics to gauge development and community engagement including tracking the number of active smart contracts on the network, monitoring the development and usage of decentralized applications, evaluating contributions to the codebase, including commits and pull requests, categorizing developers as full-time, part-time, or one-time based on their GitHub activity. Full-time developers contribute on 10 or more days per month, part-time developers contribute on fewer than 10 days, and one-time developers contribute just once, measuring participation in forums, social media, and other community platforms. These metrics are analyzed through the following parameters and aid in understanding the health and growth of the XDC Network's ecosystem as shown in Table 6. The figures indicate the real-time values as on 19-07-2024 at 15:08 IST [17].

Table 6  GitHub Statistics

| Parameter | Description | Count |
|---|---|---|
| Stars | The popularity and community support, with higher stars often indicating a more reputable and well-maintained project. | 60 for XinFin-Node |
| Forks | Level of collaborative interest, with a higher number indicating an active developer community. | 20K for XDPoSChain |
| Open Issues | Project's responsiveness to community feedback and potential bugs, with fewer open issues indicating a more stable project. | 9 for XDPoSChain |
| Watchers | The number of users following a repository for updates reflects the level of ongoing interest and engagement with the project. | 10 for XDPoSChain |
| Repository Size | Project complexity and maturity, with larger repositories containing more features but potentially requiring more maintenance. | Total Repositories on GitHub: 323; Total XDC Commits: 268,628; Monthly Active Devs: 30; Total Full-Time Devs:10 [18] |
| Network Count | The number of repositories that are forks of the original repository, providing another angle on the project's influence and reach within the developer community. | 100s for XDPoSChain |
| Subscribers | The number of users subscribed to receive notifications for updates can indicate the level of active engagement and interest in the project's future developments. | Github: 1200 subscribers; Twitter account: 100.4K followers [19]. But now the account is suspended; Reditt: 10 K subscribers [20]. |

# 5  Business

This section covers two key business factors, namely transaction cost and predictability. The average fee paid to execute a transaction on a network is referred to as cost. Users tend to find lower costs more appealing, but speed and network security must be weighed against each other. The predictability of transaction costs over a specified period is measured. Increased constituent utilization and improved usability of the network can be achieved with a more stable cost structure. Because of its architecture and consensus process, the XDC Network provides excellent predictability in terms of costs and transaction times [9]. The following parameters play important roles in the assessment the XDC network with regard to business aspects. The figures indicate real-time values as on 19-07-2024 at 15:08 IST.

- *Low and Stable Fees:* The transaction cost on the XDC Network is extremely low, approximately $0.00040485 per transaction. This low fee structure remains stable, making it predictable for users and enterprises to estimate their transaction costs.
- *High Throughput and Fast Transaction Times:* The XDC Network is capable of processing on an average over 13.30 transactions per second (TPS). This high throughput ensures that transactions are processed quickly and efficiently. The average block time on the XDC Network is around 2.0099 seconds, and transactions typically reach finality within 30 seconds. This rapid processing time ensures that users can predict when their transactions will be confirmed and settled.
- *Consensus Mechanism:* The XDC Network uses XDPoS consensus mechanism. This mechanism involves validators performing a small amount of Proof of Work (PoW) when proposing a new block to achieve consensus [21]. XDPoS is designed to be efficient, decentralized, and flexible, leveraging the power of stakeholders to resolve consensus issues fairly and democratically. This method ensures the network consumes minimal energy and is highly resistant to spamming and thereby contributes to the network's efficiency and security. This consensus model helps maintain consistent transaction processing times and reduces the likelihood of network congestion.
- *Scalability:* Even in times of heavy demand, transaction times and costs can be predicted because to the XDC Network's architecture, which can manage high transaction volumes without experiencing appreciable delays or fee hikes.
- *Reward Mechanism:* An epoch, or 900-block iteration, has a checkpoint block constructed just for the purpose of implementing reward processes. All block signatures are stored in this block, also known as the block signer. The number of signatures submitted to the smart contract by block signers during the epoch is what counts toward incentives; the more signatures linked to a Masternode, the higher the payout.
- *Fee Structure Transparency:* The fee structure of the network is open, and blockchain explorers and other network tools make it simple for users to find out about the costs and processing times of transactions. Users may more easily plan and anticipate their transaction costs and timings because of this openness with a Gas price of 250 mwei, and maximum block size of $42 \times 10^7$ gas.

The XDC Network offers very predictable transaction costs and timeframes because to its mix of low fees, fast throughput, quick transaction finality, and a strong consensus process. For enterprise applications and other use cases that demand continuous performance and cost-efficiency, this predictability is especially helpful.

## 6 Open Issues and Challenges

Like many blockchain projects, the XDC Network remains challenged with several issues related to community metrics and development. Even though the XDC Network seeks to address scalability concerns, ongoing enhancements are required to effectively manage rising transaction volumes. Maintaining smooth communication with other blockchain networks and conventional banking institutions is still a top priority. There is constant work to expand the XDC Network's user base and increase developers' and businesses' knowledge of its advantages. The XDC Network needs developers to come and stay for it to expand. This involves offering reliable development tools and assistance. It can be complicated to navigate the complicated regulatory environment surrounding blockchain technology and to ensure compliance with different countries.

## 7   Conclusion

Every facet of network assessment, such as the XDC network's security, scalability, and decentralization, has been covered in this study. The high real-time computation of the Nakamoto Coefficient in the XDC network illustrates its extremely decentralized character. The XDC Network's scalability is mainly based on its capacity to process large numbers of transactions per second (TPS) at a speed of 2,000 transactions per second through the use of XDPoS, which guarantees fast transaction speeds and low energy usage. Global financial transactions and enterprise-level applications require this high throughput. Its scalability is further enhanced by the near-zero gas fees that enable carrying out a large number of transactions economically. The network's compatibility with the Ethereum Virtual Machine (EVM) and support for interoperable smart contracts increase its scalability. The XDC Network's hybrid architecture promotes spontaneous block finality and interoperability, which supports scalability and high speed. Since the launch of the Mainnet launch 2019, XDC Network has prioritized security as part of its purpose. It is an enterprise-grade Layer 1 blockchain that is accessible to anyone, supporting a wide range of distinctive decentralized apps and providing strong security features to safeguard users' assets and information. By properly electing masternodes through an incentive program, the XDPoS consensus mechanism, which is at the center of the XDC network, governs the XDC nodes in maintaining the consistency of a decentralized ledger (the blockchain) with high security and performance guarantees. To preserve a robust and safe ecosystem, execution nodes are also incentivized to join and maintain the XDC network. XDC promises to further enhance security and scalability in the near future by increasing the standby nodes. Consequently, the XDC network makes sure that its users can transact with confidence by applying powerful consensus processes and placing a high priority on privacy and security.